\documentclass[twocolumn]{revtex4}
\usepackage{amssymb,amsmath}
\usepackage{psfrag,verbatim}
\usepackage[dvips]{graphicx}
\usepackage{color}
\definecolor{darkblue}{rgb}{0,0,0.5}
\usepackage[dvips, unicode, colorlinks, bookmarks=true, citecolor=darkblue, linkcolor=darkblue, urlcolor=darkblue]{hyperref}

\begin{document}

\begin{abstract}

The future laser interferometric gravitational-wave detectors sensitivity can be improved using squeezed light. In particular, recently a scheme which uses the optical field with frequency dependent squeeze factor, prepared by means of a relatively short ($\sim30\,{\rm m}$) amplitude filter cavity, was proposed \cite{Corbitt2004-3}. Here we consider an improved version of this scheme, which allows to further reduce the quantum noise by exploiting the quantum entanglement between the optical fields at the filter cavity two ports. 

\end{abstract}

\title{Increasing future gravitational-wave detectors sensitivity by means of amplitude filter cavities and quantum entanglement}

\author{F.Ya.Khalili}

\email{farid@hbar.phys.msu.ru}

\affiliation{Physics Faculty, Moscow State University, Moscow 119992, Russia}


\maketitle


\section{Introduction}

In high-sensitive optical position meters, and, in particular, in laser interferometric gravitational-wave detectors, two kinds of optical quantum noises exist. The first one --- {\em measurement noise}, known also as {\em shot noise} --- arises due to the optical field phase fluctuation. Its spectral density $S_x$ is inversely proportional to the optical power $W$ circulating in the interferometer arms. The second one --- {\em back action}, or {\em radiation-pressure} noise --- is a random force acting on test mass(es) due to the optical field amplitude fluctuations. This noise spectral density $S_F$ is directly proportional to $W$. 

In the contemporary (first) generation of laser interferometric gravitational-wave detectors \cite{Abramovici1992, Ando2001, Willke2002, LIGOsite, GEOsite, TAMAsite}, the optical power is relatively small and the measurement noise dominates in all frequency band of interest. In the planned second generation detectors \cite{Thorne2000, Fritschel2002, AdvLIGOsite, LCGTsite} the circulating power will be about hundred times higher. Correspondingly, sensitivity will be better at least by one order of magnitude. Near the characteristic frequency $\Omega_0/2\pi\sim100\,{\rm Hz}$ it will be close to the Standard Quantum Limit (SQL), {\em i.e.} the sensitivity level where the test mass response on the back-action force becomes equal to the measurement noise.

It is well known that performance of laser interferometric gravitational-wave detectors can be  improved by using squeezed  quantum states of the optical field. Preparation of optical field quantum states squeezed in gravitational-wave band (10-10000 Hz) presents a significant technical problem. However, now this problem can be considered as solved \cite{McKenzie2004, Vahlbruch2006}.

Squeezed state inside the interferometer can be created by injection squeezed vacuum into the interferometer dark port. In this case, the shot noise will be suppressed by the price of increased radiation pressure noise, allowing to reach the SQL using less optical power \cite{Caves1981} (it has to be noted that the circulating power close to 1\,MW is planned for the second generation detectors). Using a frequency-dependent squeezing angle \cite{Unruh1982}, it is possible to obtain sensitivity better than the SQL. The necessary dependence can be created, in particular, by reflecting squeezed vacuum (before its injection into the interferometer) from additional detuned filter cavities \cite{02a1KiLeMaThVy, Harms2003, Buonanno2004}. However, this technology is very sensitive to the filter cavities optical losses. Therefore, in order to obtain a significant sensitivity gain, very long (kilometer-scale) filter cavities are required. 

An alternative method based on a different type of filter cavity was proposed in the article \cite{Corbitt2004-3}. In this method (which will be referred by below as CMW), the filter cavity works as a high-pass filter for the squeezed light, creating light which is squeezed only at high side-band frequencies ($\gtrsim100\,{\rm Hz}$). In this case, the measurement noise will be reduced at high frequencies only, that is in the frequency domain where it dominates. At the same time, the back action noise will remain unchanged at low frequencies (where it dominates), and will increase only at high frequencies, where its influence is negligible in comparison with the measurement noise. As a result, this method improves sensitivity at high frequencies, but does not affect it at low frequencies. 

It is important, that in the CMW procedure squeezed light does not enter into the filter cavity at high side-band frequencies, and at low side-band frequencies it simply passes through the filter cavity and not used in the measurement. Therefore, the filter cavity losses could affect the sensitivity only in the relatively narrow intermediate frequencies area, and it is possible to expect, that the CMW scheme will be less sensitive to the filter cavity losses.

Estimates, presented in the article \cite{Corbitt2004-3}, shows, however, that this method performance is not as good as it could be expected, and there is a noticeable sensitivity degradation at medium frequencies --- close to the filter cavity half-bandwidth $\gamma_f$ (see, for example, Fig.\,3 of \cite{Corbitt2004-3} and Fig.\,\ref{fig:plot1} of the current paper). This degradation stems from the fact that there is a quantum entanglement between output optical fields of two filter cavity ports \footnote{In the article \cite{Corbitt2004-3}, only one filter cavity mirror is considered as partly transparent. Therefore, the filter cavity has one physical port only, and the second one in virtual (coupled with the heat bath). However, configuration with two partly transparent mirrors, {\em i.e.} with two physical ports is also possible, see below.}. It vanishes at low and high frequencies, where the input field simply passes through or reflects from the cavity, and reaches the maximum at frequencies around $\gamma_f$. It is well known that the quantum entanglement represent an informational resource. This resource is not used in the original CMW scheme. As a result, the  redundant quantum noise arises.

Thus disadvantage was recognized in the article \cite{Corbitt2004-3}, and in order to reduce it, multiple filter cavities configurations was proposed. Here we consider another more simple and effective method, see Fig.\,\ref{fig:scheme}. The scheme proposed here  differs from the original one (see Fig.\,1 of \cite{Corbitt2004-3}) by the additional homodyne detector {\sf AHD} attached to the filter cavity idle port. Due to the quantum entanglement, output signal of this detector contains information about both the shot and the radiation pressure noises. Therefore, adding it to the output of the main homodyne detector {\sf MHD} with an optimal weight function, it is possible to remove the redundant noise, substantially improving the sensitivity at medium frequencies.

In Sec.\,\ref{sec:noises}, quantum noises of this scheme is calculated and it is shown that the additional detector allows to reach the fundamental minimum of this noise. In Sec.\,\ref{sec:sensitivity}, the scheme sensitivity is estimated for different values of the optical power and the interferometer bandwidth.

For simplicity, we suppose here that the interferometer is tuned in resonance, {\em i.e.}, the ``optical springs'' technology \cite{Buonanno2001, Buonanno2002} is not used. We suppose also, that there are no optical losses in the main interferometer itself, and take into account only losses in the filter cavity. This approximation is justified by the fact that contemporary and planned gravitational-wave detectors have kilometer-scale arm cavities, while the filter cavity length in the setup considered in article \cite{Corbitt2004-3} and here is supposed to be about a few tens of meters (it is well known that the optical losses influence is inversely proportional to the cavity length). We take into account, however, non-unity quantum efficiency of the photodetectors, which significantly cripples performance of the squeezed states based schemes. 

The main notations and parameters values used in this paper are listed in Table\,\ref{tab:notations}. For the consistency with the quantum theory of measurements, the ``double-sided'' definition of the noises spectral densities used in this paper. However, for the equivalent strain noise spectral density $S^h(\Omega)$, the ``single-sided'' definition  standard for the gravitational-wave community is used [see Eq.\,(\ref{S^h})].

\begin{table*}[t]
  \begin{tabular}{|c|c|l|}
    \hline
      Quantity    & Value for estimates                 & Description \\
    \hline
      $\Omega$    &                                     & Gravitational-wave frequency \\
      $c$         & $3\times10^8\,{\rm m/s}$            & Speed of light \\
      $\omega_p$  & $1.77\times10^{15}\,{\rm s}^{-1}$   & Optical pump frequency \\
      $m$         & $40\,{\rm kg}$                      & Test mass \\
      $L$         & $4\,{\rm km}$                    & Interferometer arms length \\
      $W$         &                                     & Power circulating in each of the arms \\
      $W_0$       & $840\,{\rm kW}$                     & Power planned for the Advanced LIGO \\
      $J=\dfrac{8\omega_pW}{McL}$ &                     & \\ [2ex]
      $J_0=\dfrac{8\omega_pW_0}{McL}$ & $(2\pi\times100)^3\,{\rm s}^{-3}$ & \\ 
      $\gamma$    &                                     & Interferometer half-bandwidth \\
      $\gamma_0=J_0^{1/3}$ & $2\pi\times100\,{\rm s}^{-1}$ & 
        ``Conventional'' interferometer half-bandwidth \\
      $l_f$       & $30\,{\rm m}$                       & Filter cavity length \\
      $e^r$       & $\sqrt{10}$                         & Input field squeezing factor \\
      $A_f^2$     & $10^{-5}$                           & Filter cavity losses per bounce \\
      $T_I^2$     &                                     & Filter cavity input mirror transmittance\\
      $T_E^2=T_I^2-A_f^2$ &                             & Filter cavity end mirror transmittance
        \\[1ex]
      $\gamma_f=\dfrac{cT_I^2}{2l_f}$ &                 & Filter cavity half-bandwidth \\[2ex]
      $\eta_f = \dfrac{T_E^2}{T_I^2}$ &                 & Filter cavity quantum efficiency \\
      $\eta$      & 0.9                                 & Photodetectors quantum efficiency \\
      $\zeta$     &                                     & Homodyne angle of {\sf AHD} \\
    \hline
  \end{tabular}
  \caption{Main notations used in this paper.}\label{tab:notations}
\end{table*}

\section{Quantum noises calculation}\label{sec:noises}

\subsection{Filter cavity}

\begin{figure}
  \includegraphics[width=0.48\textwidth]{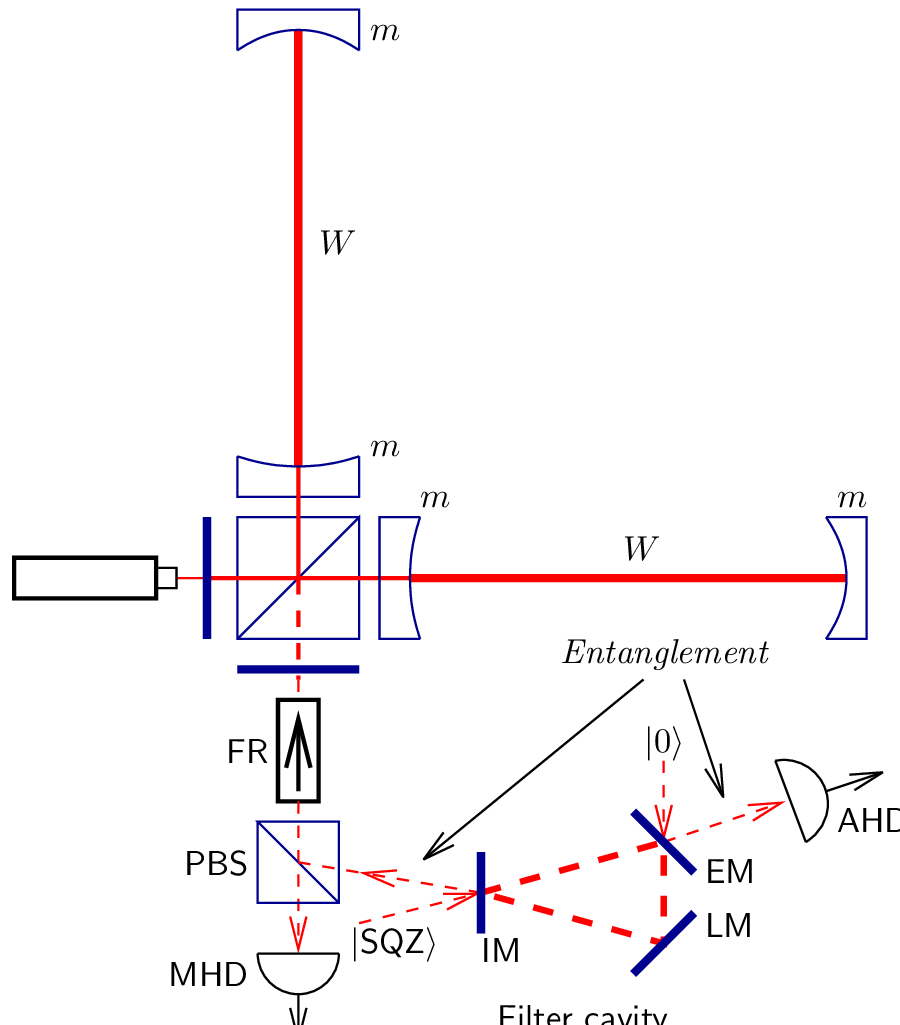}
  \caption{Squeezed vacuum is reflected from the resonance-tuned filter cavity and is fed into the gravitational-wave detector dark port using the polarization beam splitter {\sf PBS} and the Faraday rotator {\sf FR}. {\sf MHD} --- the main homodyne detector; {\sf AHD} --- the additional homodyne detector.}\label{fig:scheme}
\end{figure}

Following the article \cite{Corbitt2004-3}, consider a resonance-tuned filter cavity pumped through its input mirror {\sf IM} by squeezed vacuum (see Fig.\,\ref{fig:scheme}). Suppose that the cavity end mirror {\sf EM} is also partly transparent, and the following condition is satisfied for the mirrors power transmittance and the cavity losses per bounce:
\begin{equation}\label{filter_cond} T_I^2 = T_E^2 + A_f^2 \,. \end{equation}
The cavity tuned in such a way is transparent at low sideband frequencies $\Omega\ll\gamma_f$ and have reflectivity close to unity at high frequencies $\Omega\gg\gamma_f$. Therefore, the reflected beam which is fed into the gravitational wave detector dark port will be squeezed only at high frequencies. At low frequencies the squeezed state will be replaced by the vacuum state created  partly by the end mirror transmittance and partly by the cavity optical losses. 

It has to be noted that it was supposed in the article \cite{Corbitt2004-3} that only one (input) mirror of the filter cavity is partly transparent: $T_E=0$ and $T_I=A_f$. However, using modern  high-reflective mirrors, it is possible to obtain $A_f\ll T_I$ and, therefore, $T_E\approx T_I$. Really, the filter cavity half-bandwidth has to be close to $\gamma_f\sim2\pi\times100\,{\rm s}^{-1}$. Therefore, there has to be $T_I^2 = 2l_f\gamma_f/c\sim10^{-4}$. Taking into account that, with the best mirrors available now, the filter cavity losses per bounce can be as small as $A_f^2\approx10^{-5}$, it is possible to provide the value of $1-\eta_f = 1-T_E^2/T_I^2 = A_f^2/T_I^2\sim0.1$\,.

It is shown in Appendix \ref{app:filter}, that squeezing factors for the phase and amplitude quadrature components of the filter cavity output beam are equal to
\begin{subequations}\label{S_a}
  \begin{align}
    S_{\varphi,{\sf I}}(\Omega) &= \frac{\Omega^2e^{-2r} + \gamma_f^2}{\Omega^2 + \gamma_f^2}\,,\\
    S_{A,{\sf I}}(\Omega) &= \frac{\Omega^2e^{2r} + \gamma_f^2}{\Omega^2 + \gamma_f^2} \,,
 \end{align}
\end{subequations}
and these noises are non-correlated:
\begin{equation}\label{S_a_nc}
  S_{\varphi A,{\sf I}}(\Omega) = 0 \,.
\end{equation}
It is easy to see, that
\begin{equation}\label{product_org}
  S_{\varphi,{\sf I}}(\Omega)S_{A,{\sf I}}(\Omega)
    = 1 + \frac{4\Omega^2\gamma_f^2\sinh^2r}{(\Omega^2 + \gamma_f^2)^2} > 1 \,.
\end{equation}
It follows from Eqs.\,(\ref{S_a_nc}, \ref{product_org}), that there is a redundant quantum noise in the filter cavity output beam which is maximal at $\Omega=\gamma_f$ and vanishes if $\Omega\to0$ or $\Omega\to\infty$, {\em i.e.}, if the filter cavity is completely reflective or completely transparent. 

Consider the beam which leaves the filter cavity through the end mirror {\sf EM}. It is squeezed at low frequencies $\Omega<\gamma_f$. Squeezing factors of the corresponding phase and amplitude  quadrature components are equal to (see Appendix \ref{app:filter})
\begin{subequations}\label{S_q}
  \begin{align}
    S_{\varphi,{\sf E}}(\Omega) &= 1 - \frac{\eta_f\gamma_f^2(1-e^{-2r})}{\Omega^2+\gamma_f^2} \,,\\
    S_{A,{\sf E}}(\Omega) &= 1 + \frac{\eta_f\gamma_f^2(e^{2r}-1)}{\Omega^2+\gamma_f^2}\,.
  \end{align}
\end{subequations}
and these quadratures are also non-correlated, $S_{\varphi A,{\sf E}}(\Omega) = 0$. 

Redundant noise exists at this output too:
\begin{multline}\label{product_aux}
  S_{\varphi,{\sf E}}(\Omega)S_{A,{\sf E}}(\Omega) \\
    = 1 + \frac{4\eta_f\gamma_f^2[\Omega^2 + (1-\eta_f)\gamma_f^2]\sinh^2r}
            {(\Omega^2 + \gamma_f^2)^2} 
      > 1 \,.
\end{multline}
At the same time, there is cross-correlation between the corresponding (phase and amplitude) quadrature components at the filter cavity two outputs:
\begin{subequations}\label{IE_corr}
  \begin{align}
    S_{\varphi,{\sf IE}}(\Omega) 
      &= -\frac{i\sqrt{\eta_f}\,\Omega\gamma_f(1-e^{-2r})}{\Omega^2+\gamma_f^2} \,, \\
    S_{A,{\sf IE}}(\Omega) &= \frac{i\sqrt{\eta_f}\,\Omega\gamma_f(e^{2r}-1)}{\Omega^2+\gamma_f^2}
      \,.
  \end{align}
\end{subequations}

Equations (\ref{product_org}, \ref{product_aux}, \ref{IE_corr}) reflect a well-known feature of entangled bipartite quantum systems. Even if the whole system is in a pure quantum state, each of the parties detected individually appear as being in a mixed state. In the particular case of the optical fields at the filter cavity two outputs, this leads to the inequalities (\ref{product_org}, \ref{product_aux}). At the same time, measurement of each of the parties project the another one into a pure state, utilizing the mutual information [in our case, the cross-correlation (\ref{IE_corr})] which exists in the system due to the entanglement. 

In the particular case considered here, measurement of the optical field leaving the filter cavity though the end mirror, allows to reduce effective fluctuations of the optical field reflected from the filter cavity input mirror, compare Eqs.\,(\ref{SxSF_prod_org}) and (\ref{SxSF_prod}) below.

\subsection{Meter noise}

The component of output signal of the main homodyne detector ({\sf MHD} in Fig.\,\ref{fig:scheme}) created by the optical quantum noise is proportional to 
\begin{equation}\label{x_sum_org}
  \hat x_{\rm sum}(\Omega) = \hat x_{\rm fl}(\Omega) - \frac{\hat F_{\rm fl}(\Omega)}{m\Omega^2} \,,
\end{equation}
where $x_{\rm fl}(\Omega)$ is is the measurement noise and $\hat F_{\rm fl}(\Omega)$ is the back-action force, see Eqs.\,(\ref{CMWs_x_fl}, \ref{CMWs_F_fl}). Spectral density of the noise (\ref{x_sum_org}) is equal to:
\begin{equation}\label{S_sum_org}
  S_{\rm sum}(\Omega) = S_x(\Omega) + \frac{S_F(\Omega)}{m^2\Omega^4}\,,
\end{equation}
where
\begin{subequations}\label{S_x_S_F_org}
  \begin{gather}
    S_x(\Omega) = \frac{\hbar(\gamma^2+\Omega^2)}{4mJ\gamma}
      \left[S_{\varphi,{\sf I}}(\Omega) + \frac{1-\eta}{\eta}\right] \label{S_x_S_F_org(a)} \,, \\
    S_F(\Omega) = \frac{\hbar mJ\gamma}{\Omega^2+\gamma^2}\,S_{A,{\sf I}}(\Omega)
  \end{gather}
\end{subequations}
are spectral densities of the noises $\hat x_{\rm fl}(\Omega)$ and $\hat F_{\rm fl}(\Omega)$, correspondingly, see Appendix \ref{app:meter}. The second term in the square brackets in Eq.\,(\ref{S_x_S_F_org(a)}) arises due to the homodyne detector non-unity quantum efficiency $\eta<1$.

Due to Heisenberg's uncertainty relation, spectral densities (\ref{S_x_S_F_org}) satisfy the following inequality \cite{92BookBrKh, 03a1BrGoKhMaThVy}: 
\begin{equation}
  S_x(\Omega)S_F(\Omega) - |S_{xF}(\Omega)|^2 \ge \frac{\hbar^2}{4}\,,
\end{equation}
where $S_{xF}$ is cross-correlation spectral density, which is equal to zero in this particular case. Without optical losses or other non-fundamental noise sources, there has to be exact equality in this formula. However, it follows from Eq.\,(\ref{product_org}) that even in the case of $\eta=1$, product of spectral densities (\ref{S_x_S_F_org}) is equal to
\begin{equation}\label{SxSF_prod_org}
  S_x(\Omega)S_F(\Omega) = \frac{\hbar^2}{4}\,
    \left[1 + \frac{4\Omega^2\gamma_f^2\sinh^2r}{(\Omega^2 + \gamma_f^2)^2}\right] 
    > \frac{\hbar^2}{4} \,.
\end{equation}
Therefore, this meter is not optimal from the quantum measurements theory point of view. 
The origin of this fact is, evidently, the information loss in the filter cavity. Part of this loss created by the light absorption on the filter cavity is inevitable. But the other part created by the information leak through the filter cavity end mirror can be eliminated using the correlation (\ref{IE_corr}).

Really, suppose that some quadrature component $\hat{\rm q}_\zeta$ of the end mirror output beam, defined by the homodyne angle $\zeta$ [see Eq.\,(\ref{q_zeta})], is monitored using the additional homodyne detector {\sf AHD}. Due to the correlation (\ref{IE_corr}), this detector provides some information about the quantum fluctuations (\ref{x_sum_org}). Therefore, mixing in optimal way the main detector output with the additional detector output, it is possible to reduce the sum quantum noise. It is shown in Appendix \ref{app:adddetect}, that this optimized sum noise spectral density is equal to
\begin{equation}\label{S_sum}
  S_{\rm sum}^{\rm eff}(\Omega) 
  = S_x^{\rm eff}(\Omega) - \frac{2S_{xF}^{\rm eff}(\Omega)}{m\Omega^2} 
    + \frac{S_F^{\rm eff}(\Omega)}{m^2\Omega^4} \,,
\end{equation}
where 
\begin{subequations}\label{eff_spectras_simple}
  \begin{align}
    S_x^{\rm eff}(\Omega) &= S_x(\Omega) - \frac{|S_{x\zeta}(\Omega)|^2}{S_\zeta(\Omega)} \,, \\
    S_F^{\rm eff}(\Omega) &= S_F(\Omega) - \frac{|S_{F\zeta}(\Omega)|^2}{S_\zeta(\Omega)} \,, \\
    S_{xF}^{\rm eff}(\Omega) &= - \frac{S_{x\zeta}^*(\Omega)S_{F\zeta}(\Omega)}{S_\zeta(\Omega)} 
  \end{align}
\end{subequations}
are spectral densities of the effective measurement noise, effective back action force and the corresponding cross spectral density [for the explicit expressions, see Eqs.\,(\ref{eff_spectras})], $S_\zeta(\Omega)$ is the spectral density of the additional detector output [see Eq.\,(\ref{S_zeta})], and $S_{x\zeta}(\Omega)$, $S_{F\zeta}(\Omega)$ are the corresponding cross spectral densities [see Eqs.\,(\ref{S_xF_zeta})]. It can be shown, that if $\eta=1$, then the spectral densities (\ref{eff_spectras_simple}) satisfy the following uncertainty relation:
\begin{multline}\label{SxSF_prod}
  S_x^{\rm eff}(\Omega)S_F^{\rm eff}(\Omega) - |S_{xF}^{\rm eff}(\Omega)|^2 \\
  = \frac{\hbar^2}{4}\left[1 
      + \frac{4\Omega^2\gamma_f^2\sinh^2r}{(\Omega^2+\gamma_f^2)^2}\,
        \frac{1-\eta_f}{S_\zeta(\Omega)} 
    \right],  
\end{multline}
where $S_\zeta(\Omega)$ is the noise ${\rm q}_\zeta$ spectral density [compare with Eq.\,(\ref{SxSF_prod_org})]. In the ideal no-losses case of $\eta_f=1$, the RHS of this equation is equal to $\hbar^2/4$, which means that the noises in this case are as small as allowed by quantum mechanics.

Spectral density (\ref{S_sum}) depends on two free parameters: $\zeta$ and $\gamma_f$, which should be tuned in some optimal way. It is evident that there is no the unique universal optimization here, and the choice of $\zeta$ and $\gamma_f$ depends on the desired shape of $S_{\rm sum}^{\rm eff}(\Omega)$. Below we will use the following simple procedure which provides smooth broadband spectral density. 

\begin{figure}
  \includegraphics[width=0.48\textwidth]{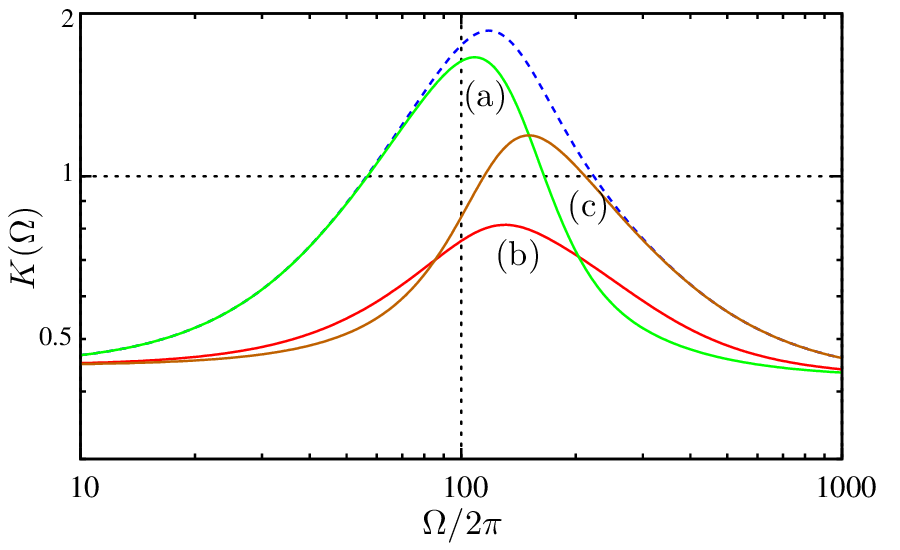}
  \caption{Typical shape of the function $K(\Omega)$ for original CMW scheme (dashes) and for CMW scheme with the additional detector (solid) and with $\zeta=0$ (a), $\zeta=\zeta_{\rm optimal}$ (b),  $\zeta=\pi/2$ (c).}\label{fig:opt}
\end{figure}

Consider the ratio of $S_{\rm sum}^{\rm eff}(\Omega)$ to the spectral density of the  corresponding ordinary (without squeezing) meter with increased by $e^r$ optical power:
\begin{equation}
  K(\Omega) = \frac{S_{\rm sum}^{\rm eff}(\Omega)\bigr|_W}
    {S_{\rm sum}^{\rm usqueezed}(\Omega)\bigr|_{We^r}}  \,.
\end{equation}
 ``Corresponding'' means that the both meters has the same values of $m$ and $\gamma$. Characteristic spectral dependences of $K(\Omega)$ is shown in Fig.\,\ref{fig:opt}. There is one maximum here close to the frequency $\Omega=\gamma_f$, and if $\Omega\ll\gamma_f$ or $\Omega\gg\gamma_f$, then $K(\Omega)\approx e^{-r}$. We will use the values of $\zeta$ and $\gamma_f$, which provide the minimal value of this maximum. 

\section{The sensitivity}\label{sec:sensitivity}

\subsection{Conventional interferometer}

In this subsection, we suppose that the interferometer half-bandwidth $\gamma$ is equal to $\gamma_0$. Below this case will be referred to as ``conventional interferometer''. 

\begin{figure}
  \includegraphics[width=0.48\textwidth]{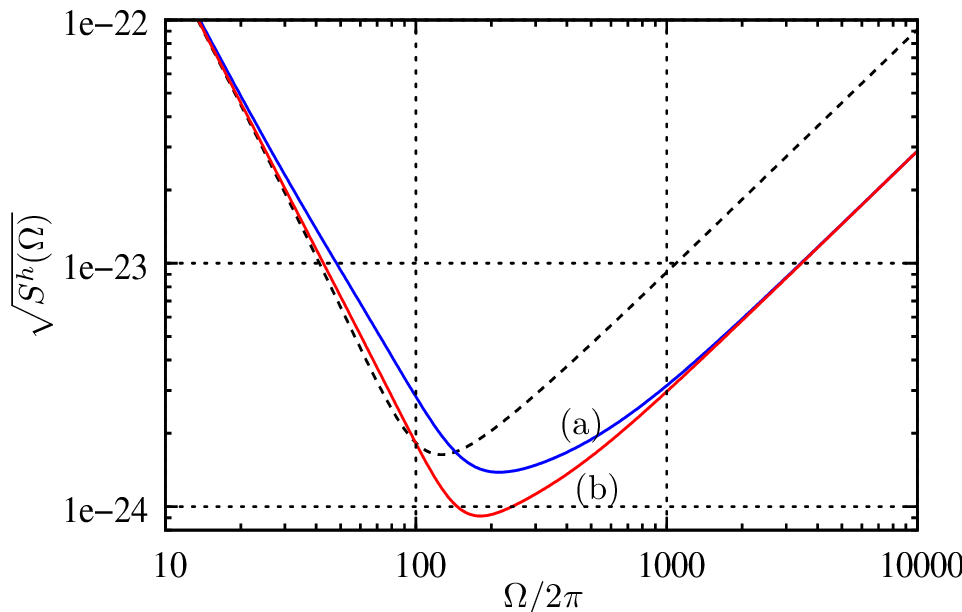}
  \caption{Square root of the sum quantum noise spectral density for: conventional unsqueezed interferometer (dashes); original CMW scheme (a); CMW scheme with the additional homodyne detector (b).  In all three cases, $W=840\,{\rm kW}$, $\gamma=2\pi\times100\,{\rm s}^{-1}$, and $\eta=1$.}\label{fig:plot1}
\end{figure}

In Fig.\,\ref{fig:plot1}, we plot the quantum noise spectral densities of the scheme with the additional detector considered above, of the original CMW scheme  and of the SQL-limited (unsqueezed) interferometer. The noises are normalized as equivalent strain fluctuations:
\begin{equation}\label{S^h}
  S^h(\Omega) = \frac{8\hbar}{mL^2\Omega^2}\,S^x(\Omega) \,.
\end{equation} 
The circulating optical power $W$ is supposed to be equal to the value planned for the Advanced LIGO ($W_0=840\,{\rm kW}$). It is easy to see that the additional detector allows to significantly reduce the noise at medium and low frequencies, compare lines (a) and (b). At low frequencies, the noise virtually does not differ from the one of the ordinary interferometer (without filter cavity).

\begin{figure}
  \includegraphics*[width=0.48\textwidth]{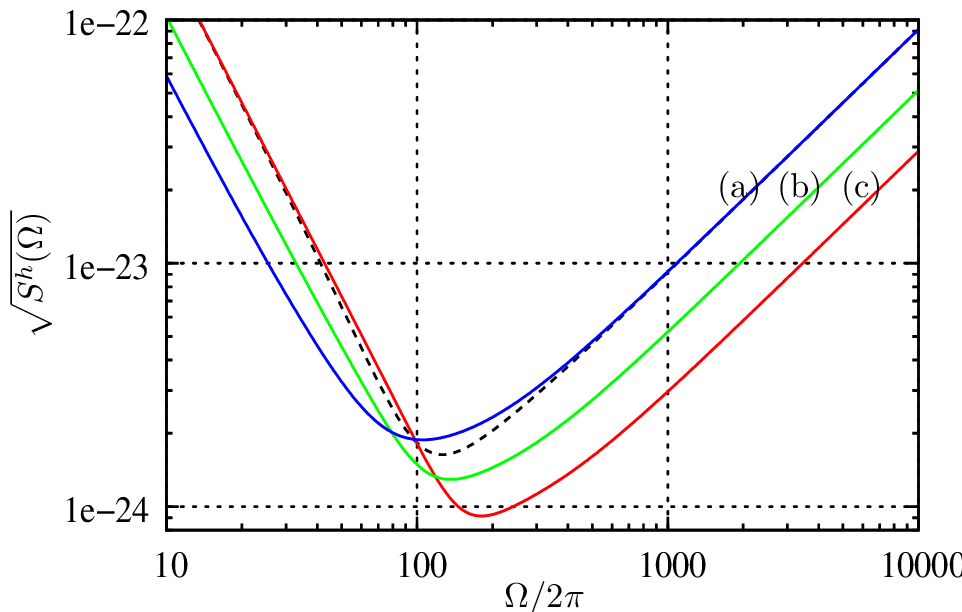}
  \caption{Square root of the sum quantum noise spectral density for CMW scheme with the additional homodyne detector for different values of the optical power: (a) $W_0e^{-2r}\approx 84\,{\rm kW}$, (b) $W_0e^{-r}\approx 266\,{\rm kW}$, (c) $W_0=840\,{\rm kW}$ (solid lines), in comparison with the conventional unsqueezed interferometer at $W=840\,{\rm kW}$ (dashes). In all cases, $\gamma=2\pi\times100\,{\rm s}^{-1}$ and $\eta=1$.}\label{fig:plot19}
\end{figure}

It is well known that reducing the optical power, it is possible to increase sensitivity at low frequencies (in the radiation-pressure noise domination area), sacrifying the high-frequency sensitivity (in the shot noise domination area). In Fig.\,\ref{fig:plot19}, the noises of the scheme considered here are plotted for the following three values of the optical power: $W_0\,e^{-2r}=84\,{\rm kW}$, $W_0\,e^{-r}=266\,{\rm kW}$, and $W_0=840\,{\rm kW}$, together with the quantum noise of the conventional unsqueezed interferometer at $W=840\,{\rm kW}$. 

The power $W=W_0\,e^{-r}$ provides almost even gain in the quantum noise spectral density in the all frequency band [see line (b)], similar to the regime with frequency-dependent homodyne angle \cite{Unruh1982,02a1KiLeMaThVy}. The gain value is smaller than in the latter case, $e^r$ vs.\! $e^{2r}$. However, the required optical power is also $e^r$ times smaller. Regime with $W=W_0\,e^{-2r}$ allows to reduce the quantum noise spectral density at low frequencies by $e^{2r}$, while keeping the high-frequency noise at the same level as in the unsqueezed interferometer. 

\begin{figure}
  \includegraphics[width=0.48\textwidth]{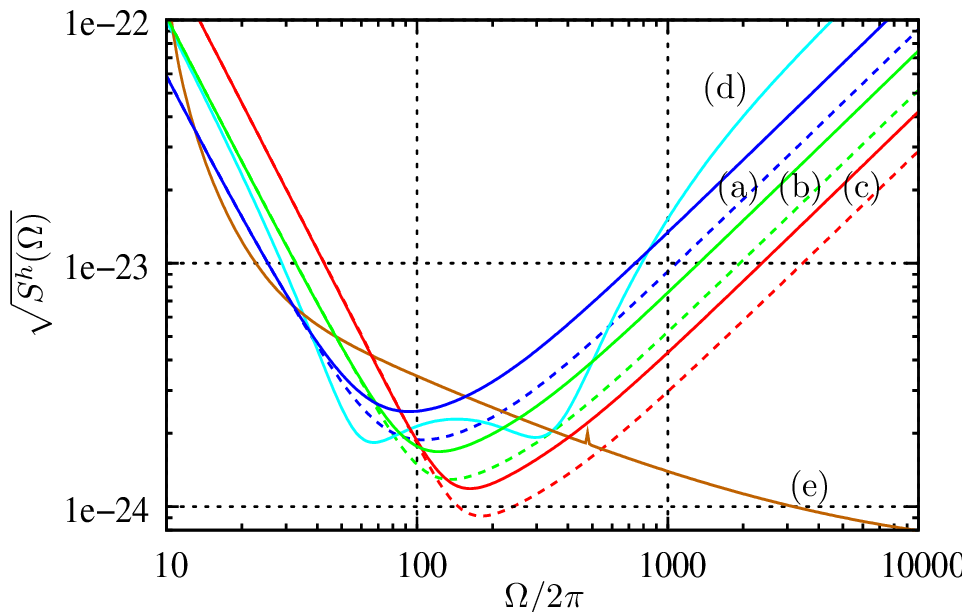}
  \caption{Square root of the sum quantum noise spectral density for CMW scheme with the additional homodyne detector for different values of the optical power: (a) $W_0e^{-2r}=84\,{\rm kW}$, (b) $W_0e^{-r}\approx 266\,{\rm kW}$, (c) $W_0=840\,{\rm kW}$. In all three cases, $\gamma=2\pi\times100\,{\rm s}^{-1}$. Solid: $\eta=0.9$, dashes: $\eta=1$. (d) The sum quantum noise of the Advanced LIGO; (e) the sum technical noise of the Advanced LIGO.}\label{fig:plot2}
\end{figure}

In Fig.\,\ref{fig:plot2}, we plot the quantum noise spectral densities for the same parameters values as in Fig.\,\ref{fig:plot19}, but taking into account non-unity quantum efficiency $\eta$ of the photodetectors (solid lines). It is easy to observe significant sensitivity degradation at high frequencies (where the light is squeezed). At the same time, the sensitivity virtually is not affected at low frequencies. For comparison, spectral densities of the sum quantum noise $S^h_{\rm quant}$ and the sum technical noise $S^h_{\rm tech}$ planned for the Advanced LIGO and calculated using the {\sc bench} program \cite{benchsite} are also plotted in Fig.\,\ref{fig:plot2}.

\begin{table*}
  \begin{center}
    \begin{tabular}{|l|c|c|c|c|c|}
      \hline
      Configuration & $W$ & $\gamma$ & ${\rm SNR}_{\rm NSNS}$ & 
        ${\rm SNR}_{\rm periodic}(1\,{\rm kHz})$ & ${\rm SNR}_{\rm periodic}(10\,{\rm kHz})$ \\
      \hline
      AdvLIGO & $840\,{\rm kW}$ &                             & 1.18      & 0.69      & 0.51 \\
      \hline
      CMW conventional & 
        $84\,{\rm kW}$  &  $2\pi\times100\,{\rm s}^{-1}$ & 1.11(1.18)& 0.69(0.99)& 0.69(1.0) \\
      & $266\,{\rm kW}$  &  $2\pi\times100\,{\rm s}^{-1}$ & 1.11(1.16)& 1.21(1.7) & 1.22(1.8)  \\
      & $840\,{\rm kW}$  &  $2\pi\times100\,{\rm s}^{-1}$ & 1.06(1.11)& 2.06(2.8) & 2.2(3.2) \\
      \hline
      CMW broadened & 
        $266\,{\rm kW}$  &  $2\pi\times314\,{\rm s}^{-1}$ & 1.22(1.29)& 2.0(2.7) & 2.2(3.2)  \\
      & $840\,{\rm kW}$  & $2\pi\times1000\,{\rm s}^{-1}$ & 1.28(1.34)& 3.9(4.8) & 6.8(9.9) \\
      \hline  
    \end{tabular}
  \end{center}
  \caption{Comparison of sensitivity of different regimes of the CMW scheme with additional homodyne detector. Low-frequency sensitivity is characterized by the parameter (\ref{SNRR_NSNS}), and the high-frequency one --- by the parameter (\ref{SNRR_periodic}). For each pair of numbers, the first one corresponds to $\eta=0.9$, and the second one (in parenthesis) --- to $\eta=1$.}\label{tab:results}
\end{table*}

In Table \ref{tab:results}, numerical estimates of the sensitivity are presented. Two criteria are used in this Table. The first one is the signal to noise ratio for neutron star - neutron star (NSNS) inspiral waveforms \cite{Buonanno2004}, normalized by the value corresponding to the conventional (unsqueezed) interferometer with $\gamma=\gamma_0$ and $W=W_0$:
\begin{equation}\label{SNRR_NSNS}
  \mathrm{SNRR}_\mathrm{NSNS} = \sqrt{
    \frac{
      \displaystyle\int_{f_c}^{f_\mathrm{ISCO}}
        \frac{f^{-7/3}}{S^h(2\pi f) + S^h_{\rm tech}(2\pi f)}\,df
    }{
      \displaystyle\int_{f_c}^{f_\mathrm{ISCO}}
        \frac{f^{-7/3}}{S^h_\mathrm{conv}(2\pi f) + S^h_{\rm tech}(2\pi f)}\,df
    }
  } \,.
\end{equation} 
Here $f_c=10\,\mathrm{Hz}$ is the gravitational-wave detector low-frequency cut-off, and $f_\mathrm{ISCO}=1570\,\mathrm{Hz}$ is the gravitational-wave frequency corresponding to the Innermost Stable Circular Orbit of a Schwarzchild black hole with mass equal to $2\times1.4$ solar masses. This criterion characterizes the low-frequency sensitivity. The second criterion is the sensitivity for high-frequency periodic sources, which is simply the reciprocal value of the noise spectral density at some given frequency. Similar to \cite{Corbitt2004-3}, we use frequencies $f=1\,{\rm kHz}$ and $f=10\,{\rm kHz}$, and normalize the noise by the value corresponding to the conventional unsqueezed interferometer with $\gamma=\gamma_0$ and $W=W_0$:
\begin{equation}\label{SNRR_periodic}
  \mathrm{SNRR}_\mathrm{periodic}(f) = \sqrt{
    \frac{S^h_\mathrm{conv}(2\pi f) + S^h_{\rm tech}(2\pi f)}{S^h(2\pi f)+S^h_{\rm tech}(2\pi f)}
  }
\end{equation}

\subsection{Broadened interferometer}

\begin{figure}
  \includegraphics[width=0.48\textwidth]{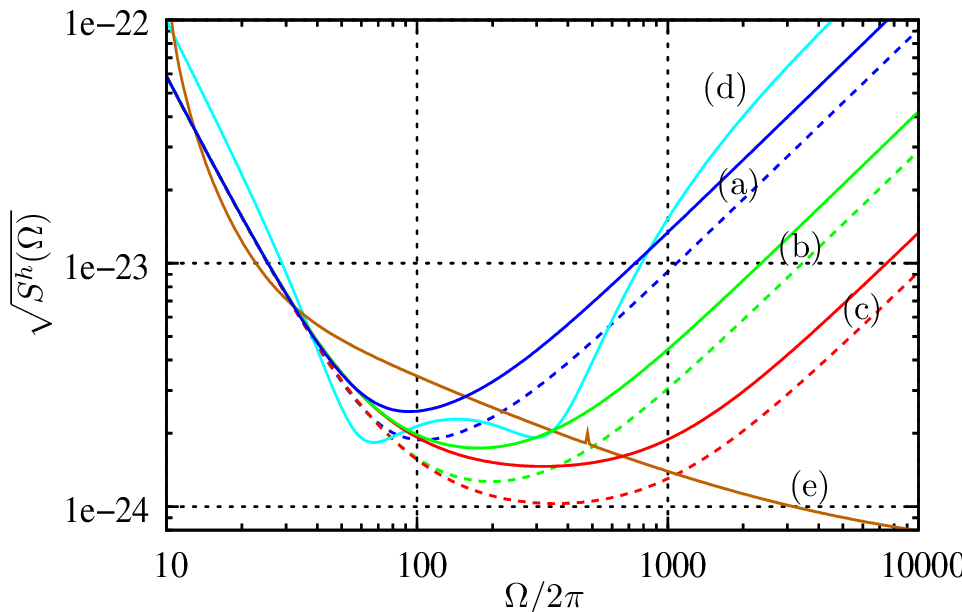}
  \caption{Square root of the sum quantum noise spectral density for CMW scheme with the additional homodyne detector for different values of the optical power and the interferometer bandwidth: (a) $W_0e^{-2r}=84\,{\rm kW},\,\gamma_0=2\pi\times100\,{\rm s}^{-1}$, (b) $W_0e^{-r}\approx 266\,{\rm kW},\,\gamma_0e^r\approx2\pi\times314\,{\rm s}^{-1}$, (c) $W_0=840\,{\rm kW},\,\gamma_0e^{2r}=2\pi\times1000\,{\rm s}^{-1}$. Solid: $\eta=0.9$, dashes: $\eta=1$. (d) The sum quantum noise of the Advanced LIGO; (e) the sum technical noise of the Advanced LIGO.}\label{fig:plot3}
\end{figure}

The choice of $\gamma=\gamma_0$, while convenient for comparison with the simplest unsqueezed SQL-limited case, does not provide the best results. Increasing not just the optical power but also the interferometer bandwidth, it is possible to improve the sensitivity at high frequencies without sacrifying the low-frequency sensitivity. Really, it follows from Eqs.\,(\ref{S_sum_org},\ref{S_sum}), that the low- and-high frequency asymptotics of the quantum noise spectral density are equal to
\begin{equation}
  S^x(\Omega) \approx \frac{\hbar}{2m}\times\begin{cases}
    \dfrac{2J}{\Omega^4\gamma} \,, & \Omega\to0 \,, \\[1.5ex]
    \dfrac{\Omega^2}{2J\gamma}\,e^{-2r} \,, & \Omega\to\infty \,.
  \end{cases}  
\end{equation}
Therefore, if $\gamma$ increases proportionally to $J\propto W$, then the low-frequency asymptotic of $S^x(\Omega)$ remains unchanged, and the high-frequency one decreases proportionally to $J^2\propto W^2$. This consideration is illustrated by Fig.\,\ref{fig:plot3}, where the quantum noise spectral density is plotted for the same values of the optical power as in Fig.\,\ref{fig:plot2}, but for proportionally increased values of $\gamma$. The corresponding numerical estimates are also shown in Table \ref{tab:results} (see the last two rows).

\section{Conclusion}

These estimates show that using the amplitude filter based scheme of \cite{Corbitt2004-3} in combination with the additional homodyne detector and the broadened interferometer configuration, it is possible to obtain sensitivity which is comparable with or better that the one planned for the Advanced LIGO at low frequencies, and substantially better at high frequencies, using $e^r$ or even $e^{2r}$ times less optical power. Taking into account that 6\,dB squeezing in gravitational waves frequency band is available now \cite{McKenzie2004, Vahlbruch2006}, and very probably 10\,dB squeezing will be available in the next few years, and taking also into account numerous hassles which stem from very high optical power planned for the next generation gravitational-wave detectors, the use of some squeezed states/amplitude filters based scheme with reduced power, similar to the one considered here, probably, could be the better option. 

However, it has to be noted that the use of squeezed states imposes additional technical problems of its own. In particular, the estimates presented in this paper show, that the photodetectors non-unity quantum efficiency $\eta<1$ substantially limits the squeezed states based scheme performance, replacing, in effect, the noise depression factor $e^{-2r}$ by the value
\begin{equation}
  e^{-2r_{\rm eff}} = e^{-2r} + \frac{1-\eta}{\eta} \,.
\end{equation}
For example, if $\eta=0.9$, then the effective squeeze factor $e^{r_{\rm eff}}$ can not exceed $\sim10\,{\rm dB}$ even if $e^{r}\to\infty$.

\acknowledgments

This work was supported by NSF and Caltech grant PHY-0353775. The paper has been assigned LIGO document number P070075-00.

The author is grateful to Yanbei Chen for usefull remarks.

\appendix

\section{Filter cavity}\label{app:filter}

\begin{figure}
  \includegraphics[width=0.48\textwidth]{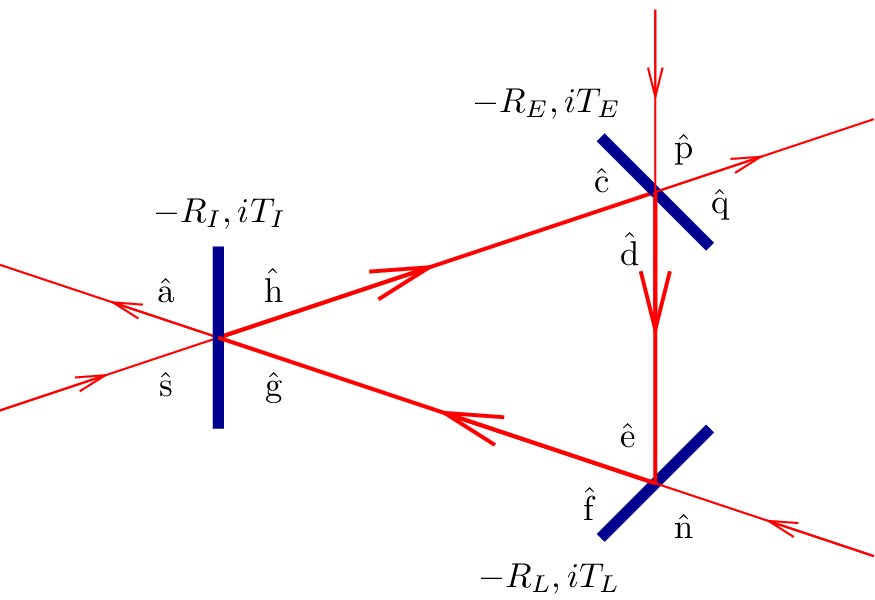}
  \caption{Lossy amplitude filter cavity}\label{fig:CMW_loss}
\end{figure}

Consider resonance-tuned ring cavity consisting of three mirrors (see Fig.\,\ref{fig:CMW_loss}): the input mirror {\sf I}, the end mirror {\sf E} and the ``loss'' mirror {\sf L}, which transmittance models the cavity optical losses. Let $-R_I,\,iT_I$, $-R_E,\,iT_E$ and $-R_L,\,iT_L$ be the reflectivities and transmittances of these mirrors, correspondingly.

Suppose that the beam  $\hat{\rm s}$ is in a squeezed state: 
\begin{equation} 
  \hat{\rm s}(\omega) = \hat{\rm z}(\omega)\cosh r - \hat{\rm z}^+(\omega)\sinh r\,,  
\end{equation} 
and the beams $\hat{\rm n}$, $\hat{\rm p}$, $\hat{\rm z}$ are in the ground state.

Equations for the field amplitudes are the following (see notations in Fig.\,\ref{fig:CMW_loss}):
\begin{subequations}\label{field_eqs}
  \begin{align}
    \hat{\rm c} &= \hat{\rm b}e^{i\omega\tau_1} \,, &
    \hat{\rm d} &= -R_E\hat{\rm c} + iT_E\hat{\rm p} \,, \\
    \hat{\rm e} &= \hat{\rm d}e^{i\omega\tau_2} \,, &
    \hat{\rm f} &= -R_L\hat{\rm e} + iT_L\hat{\rm n} \,, \\
    \hat{\rm g} &= \hat{\rm f}e^{i\omega\tau_3} \,, &
    \hat{\rm h} &= -R_I\hat{\rm g} + iT_I\hat{\rm s} \,, \\
    \hat{\rm a} &= -R_I\hat{\rm s} + iT_I\hat{\rm g} \,, &
    \hat{\rm q} &= -R_E\hat{\rm p} + iT_E\hat{\rm c} \,.
  \end{align}
\end{subequations}
where $\tau_{1,2,3}=l_{1,2,3}/c$ and $l_{1,2,3}$ are the distances between the mirrors {\sf I} and {\sf A}, {\sf A} and {\sf E}, {\sf E} and {\sf I}, correspondingly. It follows from Eqs.\,(\ref{field_eqs}), that
\begin{subequations}
  \begin{multline}
    \hat{\rm a}(\omega) = \frac{1}{1 + R_IR_ER_Le^{2i\omega\tau}}\Bigl[
      -(R_I + R_ER_Le^{2i\omega\tau})\hat{\rm s}(\omega) \\
      + R_LT_IT_E\hat{\rm p}(\omega)e^{i\omega(\tau_2+\tau_3)} 
      - T_IT_L\hat{\rm n}(\omega)e^{i\omega\tau_3}
    \Bigr] \,, 
  \end{multline}  
  \begin{multline}
    \hat{\rm q}(\omega) =  \frac{1}{1 + R_IR_ER_Le^{2i\omega\tau}}\Bigl[
      -T_IT_E\hat{\rm s}(\omega)e^{i\omega\tau_1} \\
      - (R_E + R_IR_Le^{2i\omega\tau})\hat{\rm p}(\omega)
      + R_IT_ET_L\hat{\rm n}(\omega)e^{i\omega(\tau_1+\tau_3)}
    \Bigr]\,, 
  \end{multline}
\end{subequations}
where
\begin{equation} \tau = \frac{\tau_1 + \tau_2 + \tau_3}{2} \,. \end{equation}

Suppose that the cavity is tuned in resonance:
\begin{equation}
  e^{2i\omega_p\tau} = -1 \,,
\end{equation}
and that 
\begin{align}\label{sub_exps}
  e^{i\omega_p\tau_1} = e^{i\omega\tau_3} &= 1\,, & e^{i\omega_p\tau_2} = -1
\end{align}
(conditions (\ref{sub_exps}) are not necessary, but they simplify formulae and do not affect the end results). Suppose  also that 
\begin{align}
  T_{I,E,L} &\equiv 2\sqrt{\gamma_{I,E,L}\tau} \ll 1 \,, & |\Omega|\tau \ll 1 \,,
\end{align}
where
\begin{equation}  \Omega = \omega-\omega_p \,. \end{equation}
In this case,
\begin{subequations}
  \begin{gather}
    R_{I,E,L} \approx 1 - \frac{T_{I,E,L}^2}{2} = 1 - 2\gamma_{I,E,L}\tau \,, \\
    e^{2i\omega\tau} = -e^{2i\Omega\tau} \approx -(1+2i\Omega\tau) \,,
  \end{gather}  
\end{subequations}
and
\begin{subequations}\label{CMW_1_aq}
  \begin{multline}
    \hat{\rm a}(\omega) = \frac{1}{\gamma_f-i\Omega}\Bigl[
      (\gamma_{fI}-\gamma_{fA}-\gamma_{fE}+i\Omega)\hat{\rm s}(\omega) \\
      - 2\sqrt{\gamma_{fI}\gamma_{fA}}\,\hat{\rm p}(\omega)
      - 2\sqrt{\gamma_{fI}\gamma_{fE}}\,\hat{\rm n}(\omega)
    \Bigr] \,, 
  \end{multline}
  \begin{multline}
    \hat{\rm q}(\omega) = \frac{1}{\gamma_f-i\Omega}\Bigl[
      (\gamma_{fA}-\gamma_{fI}-\gamma_{fE}+i\Omega)\hat{\rm p}(\omega) \\
      - 2\sqrt{\gamma_{fI}\gamma_{fA}}\,\hat{\rm s}(\omega) 
      + 2\sqrt{\gamma_{fE}\gamma_{fE}}\,\hat{\rm n}(\omega)
    \Bigr] \,.
  \end{multline}
\end{subequations}

Introduce new effective noises:
\begin{subequations}
  \begin{gather}
    \hat{\rm p}' = \sqrt{\eta_f}\,\hat{\rm p} + \sqrt{1-\eta_f}\,\hat{\rm n} \,, \\
    \hat{\rm n}' = \sqrt{\eta_f}\,\hat{\rm n} - \sqrt{1-\eta_f}\,\hat{\rm p} \,, 
  \end{gather}
\end{subequations}
where
\begin{align}
  \eta_f &= \frac{\gamma_{fE}}{\gamma_{fE}'} \,, & \gamma_{fE}' &= \gamma_{fE} + \gamma_{fL} \,.
\end{align}
If $\hat{\rm p}$ and $\hat{\rm n}$ correspond to two independent vacuum noises then $\hat{\rm p}'$ and $\hat{\rm n}'$ also correspond to two independent vacuum noises.

Using these new noises and renaming back for brevity
\begin{align}
  \hat{\rm p}' &\to \hat{\rm p} \,, & \hat{\rm n}' &\to \hat{\rm n} \,, & 
  \gamma_{fE}' \to \gamma_{fE} \,,
\end{align}
we obtain, that:
\begin{subequations}\label{CMW_2_aq}
  \begin{align}
    \hat{\rm a}(\omega) 
      &= {\cal R}(\Omega)\hat{\rm s}(\omega) + {\cal T}(\Omega)\hat{\rm p}(\omega) \,, \\
    \hat{\rm q}(\omega) &= \sqrt{\eta_f}
      \left[{\cal T}(\Omega)\hat{\rm s}(\omega) + {\cal Q}(\Omega)\hat{\rm p}(\omega)\right]
      + \sqrt{1-\eta_f}\,\hat{\rm n}(\omega) \,,
  \end{align}
\end{subequations}
where
\begin{equation} \eta_f = \frac{\gamma_{fE}'}{\gamma_{fE}} \,. \end{equation}
and 
\begin{subequations}
  \begin{gather}
    {\cal R}(\Omega) = \frac{\gamma_{fI} - \gamma_{fE} + i\Omega}{\gamma_f - i\Omega} \,, \\
    {\cal T}(\Omega) = -\frac{2\sqrt{\gamma_{fI}\gamma_{fE}}}{\gamma_f - i\Omega} \,, \\
    {\cal Q}(\Omega) = \frac{\gamma_{fE} - \gamma_{fI} + i\Omega}{\gamma_f - i\Omega} \,.
  \end{gather}
\end{subequations}
\begin{align}
\end{align}
Note the following symmetry conditions:
\begin{subequations}
  \begin{gather}
    |{\cal R}(\Omega)|^2 + |{\cal T}(\Omega)|^2 = |{\cal Q}(\Omega)|^2 + |{\cal T}(\Omega)|^2 = 1\,,
      \\
    {\cal R}^*(\Omega){\cal T}(\Omega) + {\cal T}^*(\Omega){\cal Q}(\Omega) = 0 \,.
  \end{gather}
\end{subequations}

Introduce two-photon quadrature amplitudes:
\begin{subequations}
  \begin{gather}
    \hat{\rm a}_\varphi(\Omega) = \hat{\rm a}(\omega_p+\Omega) + \hat{\rm a}^+(\omega_p-\Omega)\,,\\
    \hat{\rm a}_A(\Omega) = \frac{1}{i}
      \left[\hat{\rm a}(\omega_p+\Omega) - \hat{\rm a}^+(\omega_p-\Omega)\right] \,,
  \end{gather}
\end{subequations}
and similarly for all other field amplitudes. It follows from Eqs.(\ref{CMW_1_aq}), that
\begin{subequations}\label{a_pm}
  \begin{align}
    \hat{\rm a}_\varphi(\Omega) &= {\cal R}(\Omega)\hat{\rm z}_\varphi(\Omega)e^{-r} 
      + {\cal T}(\Omega)\hat{\rm p}_\varphi(\Omega) \,, \\
    \hat{\rm a}_A(\Omega) 
      &= {\cal R}(\Omega)\hat{\rm z}_A(\Omega)e^r + {\cal T}(\Omega)\hat{\rm p}_A(\Omega) \,, 
  \end{align}
\end{subequations}
\begin{subequations}\label{q_pm}
  \begin{multline}
    \hat{\rm q}_\varphi(\Omega) 
      = \sqrt{\eta_f}\,[{\cal T}(\Omega)\hat{\rm z}_\varphi(\Omega)e^{-r} 
        + {\cal Q}(\Omega)\hat{\rm p}_\varphi(\Omega)] \\
        + \sqrt{1-\eta_f}\,\hat{\rm n}_\varphi(\Omega) \,, 
  \end{multline}
  \begin{multline}
    \hat{\rm q}_A(\Omega) = \sqrt{\eta_f}\,[{\cal T}(\Omega)\hat{\rm z}_A(\Omega)e^r 
      + {\cal Q}(\Omega)\hat{\rm p}_A(\Omega)] \\
      + \sqrt{1-\eta_f}\,\hat{\rm n}_A(\Omega) \,.
  \end{multline}
\end{subequations}
Two-photon amplitudes $\hat{\rm n}_\varphi(\Omega)$, $\hat{\rm p}_\varphi(\Omega)$, $\hat{\rm z}_\varphi(\Omega)$, $\hat{\rm n}_A(\Omega)$, $\hat{\rm p}_A(\Omega)$, $\hat{\rm z}_A(\Omega)$ correspond to independent fluctuations with spectral densities equal to unity. Taking also into account Eq.\,(\ref{filter_cond}), it is easy to show that spectral densities of noises (\ref{a_pm}) are equal to (\ref{S_a}), spectral densities of noises (\ref{q_pm}) are equal to (\ref{S_q}), and 
and cross spectral densities of the noise pairs $\hat{\rm a}_\varphi(\Omega),\, \hat{\rm q}_\varphi(\Omega)$ and $\hat{\rm a}_A(\Omega),\,\hat{\rm q}_A(\Omega)$ are equal to (\ref{IE_corr}).

\section{Quantum noises of resonant-tuned interferometer}\label{app:meter}

We base the interferometer noises calculations on the ``scaling low'' theorem of the article \cite{Buonanno2003}, which maps the Michelson/Fabry-Perot topology to a single Fabry-Perot cavity with one movable mirror.

It is shown in Appendix B of the article \cite{06a1Kh}, that the output field of such a cavity is described by the following equation (it is supposed here that the cavity is resonance-tuned):
\begin{equation}
  \hat{\rm b}(\omega) = \frac{1}{\ell(\Omega)}\left[
    \ell^*(\Omega)\hat{\rm a}(\omega) + 2k_p{\rm E}\sqrt{\frac{\gamma}{\tau}}\,x(\Omega)
  \right]\,,
\end{equation}
and field inside the cavity --- by the equation 
\begin{equation}
  \hat{\rm e}(\omega) = \frac{1}{\ell(\Omega)}\left[
    i\sqrt{\frac{\gamma}{\tau}}\,\hat{\rm a}(\omega) + \frac{ik_p{\rm E}x(\Omega)}{\tau}
  \right] \,,
\end{equation}
where
\begin{align}
  \tau &= \frac{L}{c} \,, & k_p &= \frac{\omega_p}{c} \,, & \ell(\Omega) &= \gamma-i\Omega \,.
\end{align}
Finite quantum efficiency of the photodetector can be modeled by a grey filter with the transmittance $\eta$:
\begin{equation}
  \hat{\rm b}_{\rm detect}(\omega) 
  = \sqrt{\eta}\,\hat{\rm b}(\omega) + \sqrt{1-\eta}\,\hat{\rm u}(\omega) \,,
\end{equation}
where $\hat{\rm u}(\omega)$ is the corresponding introduced noise.

Using again two-photon quadratures and taking into account that 
\begin{equation}
  \ell^*(\Omega) = \ell(-\Omega) \,,
\end{equation}
we obtain, that 
\begin{subequations}
  \begin{multline}
    \hat{\rm b}_{\varphi\,{\rm detect}}(\Omega) = \frac{\sqrt{\eta}}{\ell(\Omega)}\left[
        \ell^*(\Omega)\hat{\rm a}_\varphi(\Omega) + 4k_p{\rm E}\sqrt{\frac{\gamma}{\tau}}\,x(\Omega)
      \right] \\
      + \sqrt{1-\eta}\,\hat{\rm u}_\varphi(\Omega) \label{b_detect} \,,
  \end{multline}
  \begin{equation}
    \hat{\rm e}_\varphi(\Omega) 
      = -\sqrt{\frac{\gamma}{\tau}}\,\frac{\hat{\rm a}_A(\Omega)}{\ell(\Omega)} \label{e_plus} \,.
  \end{equation}
\end{subequations}
We suppose that the moment of $t=0$ is chosen in such a way that ${\rm E} = |{\rm E}|$.

If the phase quadrature is registered, then Eq.\,(\ref{b_detect}) describes the measurement noise. It can be rewritten as follows:
\begin{equation}
  \hat{\rm b}_{\varphi\,{\rm detect}}(\Omega) 
  = \frac{4\sqrt{\eta}\,k_p{\rm E}}{\ell(\Omega)}\,\sqrt{\frac{\gamma}{\tau}}\,
    [x(\Omega) + \hat x_{\rm fl}(\Omega)] \,,
\end{equation}
where
\begin{equation}\label{CMWs_x_fl}
  \hat x_{\rm fl}(\Omega) 
  = \frac{1}{4k_pE}\sqrt{\frac{\tau}{\gamma}}\left[
      \ell^*(\Omega)\hat{\rm a}_\varphi(\Omega) 
      + \sqrt{\frac{1-\eta}{\eta}}\,\ell(\Omega)\hat{\rm u}_\varphi(\Omega) 
    \right] .
\end{equation}
is the measurement noise.

The back-action force is proportional to the quadrature (\ref{e_plus}):
\begin{equation}\label{CMWs_F_fl}
  \hat F_{\rm fl}(\Omega) = 2\hbar k_p{\rm E}\hat{\rm e}_\varphi(\Omega) 
  = -\frac{2\hbar k_p{\rm E}}{\ell(\Omega)}\,\sqrt{\frac{\gamma}{\tau}}\,\hat{\rm a}_A(\Omega) \,.
\end{equation}

Taking into account Eqs.\,(\ref{S_a}) and that $\hat{\rm u}_\varphi(\Omega)$, $\hat{\rm p}_\varphi(\Omega)$, $\hat{\rm z}_\varphi(\Omega)$, $\hat{\rm u}_A(\Omega)$, $\hat{\rm p}_A(\Omega)$, $\hat{\rm z}_A(\Omega)$ are uncorrelated noises with the spectral density equal to unity, it is easy to show, that spectral densities of the noises (\ref{CMWs_x_fl}, \ref{CMWs_F_fl}) are described by Eqs.\,(\ref{S_x_S_F_org}), and they are not correlated, $S_{xF}=0$. Using also Eqs.\,(\ref{q_pm}), it can be shown, that these noises are correlated with the additional homodyne detector output (\ref{q_zeta}), and the corresponding cross spectral densities are equal to (\ref{S_xF_zeta}).

\section{Additional detector}\label{app:adddetect}

Output signal of the additional detector is proportional to 
\begin{equation}\label{q_zeta}
  \hat{\rm q}_\zeta(\Omega) 
  = \sqrt{\eta}\,[\hat{\rm q}_\varphi(\Omega)\cos\zeta + \hat{\rm q}_A(\Omega)\sin\zeta]
    + \sqrt{1-\eta}\,\hat{\rm v}(\Omega) \,,
\end{equation}
where $\zeta$ is the corresponding homodyne phase and $\hat{\rm v}(\Omega)$ is the additional noises which arises due to the detector finite quantum efficiency $\eta<1$. It follows from Eqs.\,(\ref{S_q}), that spectral density of the noise (\ref{q_zeta}) is equal to
\begin{equation}\label{S_zeta}
  S_\zeta(\Omega) = 1 
    + \frac{\eta\eta_f\gamma_f^2(e^{2r}\sin^2\zeta + e^{-2r}\cos^2\zeta - 1)}{\Omega^2+\gamma_f^2}
    \,,
\end{equation}
and from Eqs.\,(\ref{IE_corr},\ref{CMWs_x_fl},\ref{CMWs_F_fl}) --- that cross spectral density of noises (\ref{x_sum_org}) and (\ref{q_zeta}) is equal to
\begin{equation}
  S_{\zeta,{\rm sum}}(\Omega) = S_{x\zeta}(\Omega) - \frac{S_{F\zeta}(\Omega)}{m\Omega^2} \,,
\end{equation}
where
\begin{subequations}\label{S_xF_zeta}
  \begin{align}
    S_{x\zeta}(\Omega) &= \sqrt{\frac{\hbar\eta}{4mJ\gamma}}\,
      (\gamma-i\Omega)S_{\varphi,{\sf IE}}(\Omega)\cos\zeta \,,\\
    S_{F\zeta}(\Omega) 
      &= \frac{\sqrt{\hbar mJ\gamma\eta}}{\gamma+i\Omega}\,S_{A,{\sf IE}}(\Omega)\sin\zeta
  \end{align}
\end{subequations}
are cross spectral densities of the noise pairs $\hat x_{\rm fl},\,\hat{\rm q}_\zeta$ and $\hat F_{\rm fl},\,\hat{\rm q}_\zeta$, correspondingly.

Combined noise of two homodyne detectors can be presented as follows:
\begin{equation}\label{x_sum_eff}
  \hat x_{\rm sum}^{\rm eff}(\Omega) 
  = \hat x_{\rm sum}(\Omega) - k(\Omega)\hat{\rm q}_\zeta(\Omega) \,,
\end{equation}
where $k(\Omega)$ is some factor which has to be optimized. Spectral density of the noise (\ref{x_sum_eff}) is equal to
\begin{equation}\label{S_eff_raw}
  S_{\rm sum}^{\rm eff}(\Omega) = S_{\rm sum}(\Omega) - 2\Re[k(\Omega)S_{\zeta,{\rm sum}}(\Omega)] 
    + |k(\Omega)|^2S_\zeta(\Omega) \,,
\end{equation}
The minimum of Eq.\,(\ref{S_eff_raw}) in $k(\Omega)$ corresponds to
\begin{equation}
  k(\Omega) = \frac{S_{\zeta,{\rm sum}}(\Omega)}{S_\zeta(\Omega)} \,,
\end{equation}
and is equal to
\begin{multline}\label{S_sum_app}
  S_{\rm sum}^{\rm eff}(\Omega)
  = S_{\rm sum}(\Omega) - \frac{|S_{\zeta,{\rm sum}}(\Omega)|^2}{S_\zeta(\Omega)} \\
  = S_x^{\rm eff}(\Omega) - \frac{2S_{xF}^{\rm eff}(\Omega)}{m\Omega^2} 
    + \frac{S_F^{\rm eff}(\Omega)}{m^2\Omega^4} \,,
\end{multline}
where
\begin{subequations}\label{eff_spectras}
  \begin{multline}
    S_x^{\rm eff}(\Omega) = S_x(\Omega) - \frac{|S_{x\zeta}(\Omega)|^2}{S_\zeta(\Omega)} \\
      = S_x(\Omega) - \frac{\hbar\eta\eta_f(\Omega^2+\gamma^2)}{4mJ\gamma}\,
          \frac{\Omega^2\gamma_f^2(1-e^{-2r})^2}{(\Omega^2+\gamma_f^2)^2}\,
          \frac{\cos^2\zeta}{S_\zeta(\Omega)} \,, \\
  \end{multline}
  \begin{multline}
    S_F^{\rm eff}(\Omega) = S_F(\Omega) - \frac{|S_{F\zeta}(\Omega)|^2}{S_\zeta(\Omega)} \\
      = S_F(\Omega) - \frac{\hbar mJ\gamma\eta\eta_f}{\Omega^2+\gamma^2}\,
          \frac{\Omega^2\gamma_f^2(e^{2r}-1)^2}{(\Omega^2+\gamma_f^2)^2}\,
          \frac{\sin^2\zeta}{S_\zeta(\Omega)}\,, \\
  \end{multline}  
  \begin{multline}
    S_{xF}^{\rm eff}(\Omega) = - \frac{S_{x\zeta}^*(\Omega)S_{F\zeta}(\Omega)}{S_\zeta(\Omega)} \\
      = \frac{\hbar\eta\eta_f}{2}\,
        \frac{\Omega^2\gamma_f^2(\cosh2r-1)}{(\Omega^2+\gamma_f^2)^2}\,
        \frac{\sin2\zeta}{S_\zeta(\Omega)} \,.
  \end{multline}
\end{subequations}


\end{document}